\documentclass{osa-article}

\usepackage{amsmath}
\usepackage{lineno}

\usepackage{graphicx}
\usepackage{bm}
\usepackage{mathtools}

\journal{osajournal}

\articletype{Research Article}

\newcommand\fdg{\mbox{$.\!\!^\circ$}}%

\begin{document}

\title{Two-dimensional Light Beam Shape Characterization using Interferometric Closure Amplitudes}

\author{Nithyanandan Thyagarajan,\authormark{1,2,*} Bojan Nikolic,\authormark{2} Christopher L. Carilli,\authormark{3} Laura Torino,\authormark{4} and Ubaldo Iriso,\authormark{4}}

\address{\authormark{1}Commonwealth Scientific and Industrial Research Organisation (CSIRO), Space \& Astronomy, P. O. Box 1130, Bentley, WA 6102, Australia \\
  \authormark{2}Astrophysics Group, Cavendish Laboratory, University of Cambridge, Cambridge CB3 0HE, UK\\
  \authormark{3}National Radio Astronomy Observatory, P. O. Box 0, Socorro, NM 87801, US\\
\authormark{4}ALBA - CELLS Synchrotron Radiation Facility, Carrer de la Llum 2-26, 08290 Cerdanyola del Vall\`es, Barcelona, Spain
}

\email{\authormark{*}Nithyanandan.Thyagarajan@csiro.au} 



\begin{abstract}
  We introduce a novel technique using closure amplitudes, inspired by radio interferometry, to determine with high angular resolution the two-dimensional profile of a light beam using an interferogram from a non-redundantly masked aperture. Previous techniques have required multiple interferograms or accurate estimates of the non-uniform illuminations across the aperture. In contrast, our method using closure amplitudes avoids the need to estimate the aperture illuminations while determining the two-dimensional beam shape from a single interferogram. The invariance of closure amplitudes to even time-varying aperture illuminations makes it suitable to longer averaging intervals, with potential to reducing data rates and computational overheads. By using data from the ALBA synchrotron light source to validate the method and its results against existing methods, this paper represents the first real-world application of closure amplitudes to directly determine the light beam's profile using optical interferometry in the high angular resolution regime.
\end{abstract}


\section{Introduction} \label{sec:intro}

High angular resolution optical imaging at the diffraction limit, typically $\lesssim 1$ arcsecond, is challenging because small imperfections in optical components, and turbulence and inhomogeneities in the medium cause fluctuations of path length resulting in ``seeing'', a loss of image quality due to decoherence. Depending on the light levels, techniques like lucky imaging, speckle masking (bispectral analysis), fully-filled aperture and non-redundant aperture masking interferometry aided by Fourier domain analysis, and adaptive optics have been used to achieve diffraction-limited high angular resolution imaging \cite{IAU158}. They all rely on preserving or restoring the coherence of light, albeit using different approaches.

Inspired by radio interferometry methods in astronomy \cite{TMS2017,SIRA-II}, recently \cite{Nikolic+2024} adapted the non-redundant aperture mask interferometry and the amplitude variant of the ``self-calibration'' concept \cite{Schwab1980,Cornwell+Wilkinson1981,Schwab+Cotton1983,Pearson+Readhead1984,TMS2017,Selfcal-SIRA} to optical interferometry at visible wavelengths to determine the two-dimensional cross-section of the light beam under high light levels. Self-calibration refers to the method of recovering the spatial coherence of the incident light beam by correcting for relative amplitude and phase variations across the masked aperture elements (holes) using a model of the light source itself. In multi-hole apertures, non-uniform illumination and phase disturbances across the holes can degrade coherence and affect the interferogram. If the hole-to-hole variations in phase and illuminations (amplitudes) are not properly accounted for, the loss of coherence will irrecoverably bias the estimated beam profile to larger sizes. \cite{Nikolic+2024} leveraged a non-redundant hole layout in the aperture plane, ensuring that each spatial frequency corresponds to a unique hole-pair. By measuring spatial correlations (visibilities) in the aperture plane, the method fits the data simultaneously for both the beam shape parameters and the hole illuminations through amplitude self-calibration. The advantage of their technique lies in its ability to determine the two-dimensional beam profile and non-uniform illumination at the unmasked holes simultaneously, provided that a sufficiently large array of non-redundantly spaced holes spans the aperture.

The above cross-disciplinary approach motivates further integration of powerful radio interferometry concepts into optical interferometry. Here, we take an alternate approach to addressing coherence loss caused by non-uniform hole illumination through the use of closure amplitudes \cite{Twiss+1960,Readhead+1980,Pearson+Readhead1984,TMS2017,SIRA-II}. Closure amplitudes form a subset of a class of interferometric data combinations, called \textit{closure invariants} \cite{Thyagarajan+2022,Samuel+2022}, that inherently eliminate the need to calibrate hole-based amplitudes that are not intrinsic to the source. Closure invariants have been instrumental in situations where calibration of the aperture behavior and propagation effects are challenging, such as in Very Long Baseline Interferometry \cite{Pearson+Readhead1984,TMS2017,VLBI-SIRA}. For example, closure invariants have played a key role in major astronomical discoveries, including the identification of morphologies and superluminal expansions of jets of quasars \cite{Jennison1958,Jennison+Latham1959,Twiss+1960,Rogers+1974,Wilkinson+1977,Pearson+1981}, and the event horizon-scale imaging of the supermassive black holes in the M87 and Milky Way galaxies \cite{eht19-1,eht21-7}.

In this paper, we introduce a novel closure amplitude-based method for optical interferometry that analytically reconstructs the two-dimensional transverse profile of the light beam and obviates the need to calibrate the hole illumination across the aperture. In a first demonstration of this framework for interferometric optical image reconstruction, we apply it to a synchrotron light source from a particle accelerator. Yet, the technique is readily generalizable to any light beam at any wavelength and not just limited to synchrotron beams.

The paper is organized as follows. The underlying motivation for the area of application and the light beam parametrization is described in section~\ref{sec:application}. It also presents the experimental setup and data used in this study. In section~\ref{sec:analysis}, we describe the framework and the reference method before presenting the novel method developed in this study. The results and conclusions are presented in sections~\ref{sec:results} and \ref{sec:conclusions}, respectively. 

\section{Application: Synchrotron Light Beam from a Particle Accelerator}\label{sec:application}

The shape of the synchrotron light beam produced from particle accelerators is closely related to a critical property of the particle beams, namely, emittance. Emittance is the area the particles occupy in the position-momentum phase space. Maintaining low emittance is a key figure of merit in particle accelerators. Low emittances enhance luminosity in colliders and maximize the flux brilliance in synchrotron light sources. Because the emittance of the particle beam is not a directly measurable parameter, it is often inferred by measuring the transverse beam shape of the synchrotron light beam coupled with the knowledge of the machine's Twiss parameters \cite{Wiedemann2015}.

Due to its non-invasive nature, imaging using the synchrotron radiation produced by the particle beam is often used to measure the transverse shape of the beam \cite{Kube2007}. For low emittance (and thus, small beam size) accelerators, direct imaging techniques use the X-ray part of the synchrotron radiation to avoid the diffraction limit (for example, the X-ray pinhole method \cite{Elleaume+1995}). At longer wavelengths, i.e. visible range, Synchrotron Radiation Interferometry (SRI) is a widely used technique.

Traditionally, SRI involved the use of a two-aperture mask at visible wavelengths, where photons passing through the unmasked hole apertures interfere on the CCD. The resulting interferogram is influenced by the size and illumination of the holes and their separations. The size of a one-dimensional beam profile can be determined by measuring how the visibility of the interferogram changes with different aperture separations.  If the beam has a Gaussian shape, this measurement can be done with a single acquisition \cite{Mitsuhashi1999,Trad+2017}. Recent advancements have enhanced SRI techniques, addressing the limitation of measuring only one-dimensional profiles. One notable development involves the use of a rotating two-hole mask at ALBA, allowing for the estimation of the two-dimensional Gaussian beam profile \cite{Torino+Iriso2016}. Another advancement employs a four-hole aperture in a square layout, capturing spatial coherence information along both vertical and horizontal directions to enable two-dimensional profiling, as demonstrated on the Spring-8 storage ring \cite{Masaki+Takano2003}. Here, we leverage a two-dimensional non-redundant aperture mask to obtain the two-dimensional cross-section of the light beam.

\subsection{Light Beam Model}\label{sec:beam-model}

The transverse profile of a synchrotron beam is well-modeled as a two-dimensional Gaussian \cite{Sands1969}, which we parametrize as 
\begin{align}
    I(\bm{\ell}) &= I_0 \, \exp\left(-\frac{1}{2} \bm{\ell}^\intercal \bm{\Sigma}_{\bm{\ell}}^{-1} \bm{\ell}\right) \, , \label{eqn:image-gaussian}
\end{align}
where, $\bm{\ell} \coloneqq \begin{bmatrix} \ell & m \end{bmatrix}^\intercal$ denotes the coordinates describing the image plane, and $\bm{\Sigma}_{\bm{\ell}}$ is the covariance matrix of a rotated two-dimensional Gaussian in this plane. The semi-principal axes of the ellipse, corresponding to the $\bm{\ell}^\intercal \bm{\Sigma}_{\bm{\ell}}^{-1} \bm{\ell} = 1$ contour, are denoted by $\sigma_\textrm{maj}$ and $\sigma_\textrm{min}$, respectively. The tilt angle of the major axis anti-clockwise from the $\ell$-axis is denoted by $\theta$. 

The spatial coherence, $\gamma(\mathbf{u})$, in the aperture plane denoted by coordinates, $\mathbf{u}\coloneqq \begin{bmatrix} u & v \end{bmatrix}^\intercal$ is related to the intensity distribution in the image plane by the van~Cittert-Zernike relation \cite{VanCittert1934} and can be approximated by a Fourier transform \cite{TMS2017,SIRA-II} in the limit of narrow field of view. Thus, for a Gaussian beam shape, 
\begin{align}
    \gamma(\mathbf{u}) &= \exp\left(-\frac{1}{2} \mathbf{u}^\intercal \bm{\Sigma}_\mathbf{u}^{-1} \mathbf{u}\right) \, , \label{eqn:aperture-gaussian}    
\end{align}
with the aperture plane covariance, $\bm{\Sigma}_\mathbf{u} = (2\pi)^{-2} \, \bm{\Sigma}_{\bm{\ell}}^{-1}$. 

The semi-principal axes of the ellipse, corresponding to $\mathbf{u}^\intercal \bm{\Sigma}_\mathbf{u}^{-1} \mathbf{u} = 1$ contour, are denoted by $\sigma_u$ and $\sigma_v$, respectively. It is convenient to further parametrize 
$\bm{\Sigma}_\mathbf{u}^{-1}=    
\begin{bmatrix}
        a & b/2 \\
        b/2 & c 
\end{bmatrix}$, 
thus reducing the equation to its quadratic form, $a\, u^2 + b\, uv + c\, v^2=1$ with $b^2-4ac<0$ (ellipse). Through an eigendecomposition of the symmetric positive-definite matrix, $\bm{\Sigma}_\mathbf{u}^{-1}=\mathbf{U} \mathbf{D} \mathbf{U}^\intercal$, with $\mathbf{U}$ holding the eigenvectors and $\mathbf{D}=\textrm{diag}(\lambda_1, \lambda_2)$ consisting of eigenvalues, $\lambda_1$ and $\lambda_2$ (with $\lambda_1 \geq \lambda_2 > 0$), the shape of the elliptical Gaussian can be expressed as 
\begin{align}
    \begin{bmatrix} \sigma_\textrm{maj} \\ \sigma_\textrm{min} \end{bmatrix} &= \frac{1}{2\pi}\begin{bmatrix} \sigma_u^{-1} \\ \sigma_v^{-1} \end{bmatrix} = \frac{1}{2\pi} \begin{bmatrix} \sqrt{\lambda_1} \\ \sqrt{\lambda_2} \end{bmatrix} \, , \label{eqn:semi-principal-axes} \\
    \theta &= \frac{1}{2} \tan^{-1}\left(\frac{b}{a-c}\right) \label{eqn:pa}
\end{align}
Alternatively, the orientation of the first eigenvector also gives the tilt angle, $\theta$, of the ellipse. 

\subsection{Experimental Setup and Data Acquisition}\label{sec:experiment-data}

\begin{figure}
  \includegraphics[width=\linewidth]
{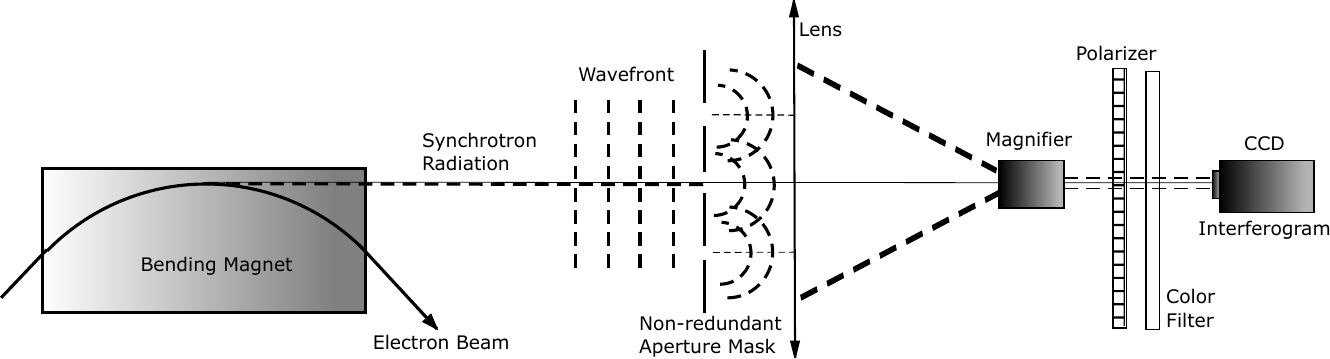}
\caption{A highly simplified sketch of the experimental setup showing the process of measuring an interferogram of the photons constituting the synchrotron radiation (visible wavelength in this case). Several optical and mechanical elements are present between the light source and the non-redundant aperture (NRA) mask that can imprint amplitude and phase distortions on the photons incident on the non-redundant aperture mask, but they are not shown here for simplicity. 
\label{fig:experiment-setup}}
\end{figure}

The experimental setup shown in Fig.~\ref{fig:experiment-setup} and the acquired data are identical to that used in \cite{Nikolic+2024}, wherein more details are provided. The details relevant for this paper are briefly described here. 

The experiments were carried out at the ALBA synchrotron facility. The visible part of the synchrotron radiation, emitted from a bending magnet, passing through a non-redundant 5-hole aperture mask, an adaptation of the non-redundant array in \cite{Gonzalez+Yobani2011}, is filtered using a $538\pm 10$~nm color filter and captured by a CCD camera. The non-redundant mask design ensures that the vector (`baseline') between every pair of apertures is unique. The motivation for this design is to avoid decoherence of the wavefront that would occur in redundant aperture measurements due to phase errors by making every spatial frequency in the Fourier domain correspond to a single pair of apertures. The use of a non-redundant 5-hole mask was to ensure that we obtained adequate independent measurements to constrain the two-dimensional Gaussian profile, as will be elaborated further in Sections~\ref{sec:VisAmp} and \ref{sec:ClAmp}.

The CCD records two-dimensional arrays of size $1296\times 966$. After removing a constant offset due to the bias and dark currents, the images are padded and centered on the Airy disk-like envelope of the interference fringes, yielding larger two-dimensional images of size $2048 \times 2048$. The aperture plane (Fourier domain) response is calculated using the Fourier transform of the padded and centered image. The visibilities on each of the baselines are calculated as the complex sum of pixels within a circular aperture of radius 7\,pixels, centered at the calculated position of the baseline. 

\section{Data Analysis in a Radio Interferometric Framework}\label{sec:analysis}

The visibility amplitude measured by a pair of holes, $p$ and $q$, in the aperture plane, when affected by the illumination of the holes at time, $t$, can be written as
\begin{align}
    |V_{pq}(t)| &= \gamma(\mathbf{u}_{pq}) \, |G_p(t)|\, |G_q(t)| \, . \label{eqn:meas-vis}
\end{align}
Mathematically, the hole illumination, $G_p(t)$, represents the amplitude scaling and phase retardation of the radiation incident at the hole, $p$, whose physical cause could have arisen from the optics and propagation effects in the traversed photon path. In radio interferometry, the complex-valued hole illumination is typically characterized as the effect on the amplitude and phase of the incident electromagnetic wave caused by instrumental and/or propagation effects due to the ionosphere or troposphere, for example. We find it applicable to SRI even though there are no measurements of amplitudes or phases of electric fields whatsoever. In this paper, we are only concerned with the amplitudes of the hole illuminations, $|G_p(t)|$.

Note that the shape of the light beam source is assumed to be constant in time, whereas the hole illuminations can be time-varying due to, for example, mechanical vibrations in the photon transport line. From the visibility amplitudes which are temporal and spatial distortions of the true amplitude of spatial coherence, we aim to derive the shape of the elliptical Gaussian beam in Eqs.~(\ref{eqn:semi-principal-axes}) and (\ref{eqn:pa}). 

First, we describe the recently developed method of amplitude self-calibration in SRI that simultaneously determines the beam shape parameters and the hole illuminations as proposed in \cite{Nikolic+2024} with a linearized variant. Then, we propose an alternate and independent method using closure amplitudes that estimates the beam shape parameters without the need for determining the hole illuminations, which is the main feature of this work. 

\subsection{Reference Method: Amplitude Self-Calibration}\label{sec:VisAmp}

The concepts of phase and amplitude self-calibration \cite{Schwab1980,Cornwell+Wilkinson1981,Schwab+Cotton1983,Pearson+Readhead1984,Selfcal-SIRA} are familiar in radio interferometry, and the amplitude variant of self-calibration was recently adapted by \cite{Nikolic+2024} for particle beam shape characterization in SRI at optical wavelengths. Usually in radio interferometry, the source model parametrization and the aperture gains are iteratively updated until convergence is achieved. In the present case, because the beam model is well-characterized by a Gaussian, the beam shape and hole illuminations (gain amplitudes) can be simultaneously estimated in a single iteration at every instant of time. 

In such a minimization process, if $N_\textrm{h}$ is the number of holes in the aperture, the number of visibility amplitude measurements at a single instant of time equals $N_\textrm{meas} = N_\textrm{h}(N_\textrm{h}-1)/2$, assuming non-redundant vector separations between the hole pairs. $N_\textrm{meas} = N_\textrm{h}(N_\textrm{h}-1)/2 + 1$ if the auto-correlation is included. The numbers of beam shape parameters and gain amplitudes that need to be determined from these measurements are $N_\textrm{shape}$ and $N_\textrm{h}$, respectively, amounting to $N_\textrm{par}=N_\textrm{shape} + N_\textrm{h}$ unknown parameters. A unique solution is possible if $N_\textrm{par} \leq N_\textrm{meas}$. In this work, $N_\textrm{h}=5$. For an elliptical Gaussian-shaped beam parametrized in Eqs.~(\ref{eqn:semi-principal-axes}) and (\ref{eqn:pa}), $N_\textrm{shape}=3$ (corresponding to $\sigma_\textrm{maj}$, $\sigma_\textrm{min}$, $\theta$). Note that if the aperture had only 4 holes, then with $N_\textrm{meas}=6$ unique visibility amplitude measurements (assuming non-redundancy), a unique solution for $N_\textrm{par}=7$ parameters ($N_\textrm{shape}=3$ parameters, $N_\textrm{h}=4$ gain amplitudes) is not possible unless the additional auto-correlation measurement is also used. Thus, $N_\textrm{h}=5$ is the minimum number of holes which will satisfy the overdetermined criterion.

We employ the amplitude self-calibration method as a reference to compare the new method we develop in Section~\ref{sec:ClAmp}. We differ slightly from \cite{Nikolic+2024} as they solved the equations using non-linear optimization that is hereafter referred to as non-linear amplitude self-calibration (NLASC). Here, we formulate a linear variant of their equations for the Gaussian coherence function by taking logarithms as:
\begin{align}
    \log |V_{pq}(t)| &= - \frac{1}{2} (a u_{pq}^2 + b u_{pq}v_{pq} + c v_{pq}^2) \nonumber \\
    &\quad  + \log |G_p(t)| + \log |G_q(t)| \, . \label{eqn:logVisAmp}
\end{align}
The equation is evidently linear in the quadratic parameters, $a$, $b$, and $c$, describing the ellipse and in the logarithms of the gain amplitudes. The measurements can be expressed as a linear equation
\begin{align}
    \mathbf{y}(t) &= \mathbf{J} \mathbf{x}(t) + \mathbf{n}(t) \, ,
\end{align}
where, 
$\mathbf{x}(t) \coloneqq \begin{bmatrix}
    a & b & c & \log |G_0(t)| & \ldots & \log |G_{N_\textrm{h}-1}(t)|
\end{bmatrix}^\intercal$, 
$\mathbf{y}(t) \coloneqq \begin{bmatrix}
    \log |V_{01}(t)| & \ldots & \log |V_{pq}(t)| & \ldots
\end{bmatrix}^\intercal$, 
$\mathbf{n}(t)$ is some random noise vector associated with $\mathbf{y}(t)$, and $\mathbf{J}$ is the \textit{Jacobian} matrix describing the measurement model that transforms $\mathbf{x}$ to $\mathbf{y}$ as specified in Eq.~(\ref{eqn:logVisAmp}). 

The advantage of this linear formulation is that the solution can be analytically determined in the form $\widetilde{\mathbf{x}}(t)=\mathbf{W}\mathbf{y}(t)$. For example, if $\mathbf{N}\coloneqq \langle\mathbf{n}(t)\,\mathbf{n}(t)^\intercal\rangle$ is the covariance of noise in the measurements, and $\bm{\varepsilon}(t)\coloneqq \widetilde{\mathbf{x}}(t) - \mathbf{x}(t)$ denotes the reconstruction error, then the solution that minimizes $\chi^2(t)\equiv (\mathbf{y}(t)-\mathbf{J}\widetilde{\mathbf{x}}(t))^\intercal \, \mathbf{N}^{-1} \, (\mathbf{y}(t)-\mathbf{J}\widetilde{\mathbf{x}}(t))$ as well as $\langle|\bm{\varepsilon}(t)|^2\rangle$ subject to $\mathbf{WJ}=\mathbf{I}$ (implying the reconstruction error, $\bm{\varepsilon}(t)$, is independent of $\mathbf{x}(t)$) is given by \cite{Tegmark1997}
\begin{align}
    \mathbf{W} &= [\mathbf{J}^\intercal\mathbf{N}^{-1}\mathbf{J}]^{-1} \mathbf{J}^\intercal \mathbf{N}^{-1} \, . \label{eqn:solution}
\end{align}
If the probability distribution of $\mathbf{n}$ is Gaussian, this solution is the maximum-likelihood estimate of $\mathbf{x}(t)$. The covariance of the errors in the solution is given by 
\begin{align}
    \langle\bm{\varepsilon(t)\,\varepsilon(t)}^\intercal\rangle &= [\mathbf{J}^\intercal\mathbf{N}^{-1}\mathbf{J}]^{-1} \, . \label{eqn:out-covariance}
\end{align} 
Thus, the shape parameters and the gain amplitudes as well as the covariance between them can be estimated analytically. 

Because the hole illuminations can be time-varying due to mechanical vibrations, the solution has to be determined in every time record, that is, as a function of time, $t$. The shape parameters determined at every instance are converted to the corresponding Gaussian shape parameters using Eqs.~(\ref{eqn:semi-principal-axes}) and (\ref{eqn:pa}), and then averaged over the accumulation interval to obtain the final estimate. We refer to this method as ``linearized'' amplitude self-calibration (LASC) to distinguish it from NLASC \cite{Nikolic+2024}.

The reference methods, LASC and NLASC, are essentially the same in concept. The main mathematical difference is that the former, on account of taking logarithms, is expressible as a linear system of equations which can be solved in an analytical framework. The other practical differences are that the LASC employs inverse-covariance weighting for optimal solution and uses only visibilities (corresponding to cross-correlations), whereas NLASC of \cite{Nikolic+2024} did not employ inverse-covariance weighting and used both auto-correlations (total intensity) and cross-correlations.

\subsection{Our Method: Direct Beam Shape Estimation using Closure Amplitudes}\label{sec:ClAmp}

Now we describe our method that is independent of the hole illumination and any errors therein, in contrast to the self-calibration approach governing the reference methods. In radio interferometry, special combinations of data that are independent of gain phases or amplitudes associable with the aperture, called \textit{closure phases} \cite{Jennison1958,Rogers+1974,Thyagarajan+Carilli2022} and \textit{closure amplitudes} \cite{Twiss+1960,Readhead+1980}, respectively, are regularly employed \cite{Pearson+Readhead1984,TMS2017,SIRA-II}. More generally, these are called \textit{closure invariants} \cite{Thyagarajan+2022,Samuel+2022}. 

As the beam shape is characterized by a Gaussian which is real-valued and has symmetry around the origin, the ideal spatial coherence is also expected to be real-valued and symmetric, specifically a Gaussian as given by Eq.~(\ref{eqn:aperture-gaussian}). Because only amplitudes are relevant for a Gaussian shape, and phases of the visibility are immaterial, the use of closure amplitudes suffices to determine the beam shape. For characterizing a more complex-shaped source, or for probing departures from Gaussianity, the use of closure phases, or phases in general, will also be required \cite{Carilli+2025}. 

The closure amplitude is usually defined on a closed loop of four aperture elements, hereafter referred to as a \textit{quad}, as \cite{Twiss+1960,Readhead+1980,TMS2017,SIRA-II}
\begin{align}
    A_{pqrs} &= \frac{|V_{pq}(t)|\,|V_{rs}(t)|}{|V_{ps}(t)|\,|V_{rq}(t)|} = \frac{|\gamma(\mathbf{u}_{pq})|\,|\gamma(\mathbf{u}_{rs})|}{|\gamma(\mathbf{u}_{ps})|\,|\gamma(\mathbf{u}_{rq})|}\, , \label{eqn:ClAmp}
\end{align}
after using Eq.~(\ref{eqn:meas-vis}). It is evident that unlike the visibility amplitude, the closure amplitude, $A_{pqrs}$, is independent of the hole illuminations, $G_p(t)$, $G_q(t)$, $G_r(t)$, and $G_s(t)$, and thus expected to remain constant in time as long as the beam shape remains stable in that interval. 

Again, we apply the logarithm on both sides to cast the equation to be linear in the Gaussian beam shape parameters,
\begin{align}
    \log A_{pqrs} &= \frac{1}{2} [(u_{ps}^2 + u_{rq}^2 - u_{pq}^2 - u_{rs}^2)\,a \nonumber\\ 
    &\quad + (u_{ps}v_{ps} + u_{rq}v_{rq} - u_{pq}v_{pq} - u_{rs}v_{rs})\, b \nonumber\\ 
    &\quad + (v_{ps}^2 + v_{rq}^2 - v_{pq}^2 - v_{rs}^2)\, c] \, . \label{eqn:logClAmp}
\end{align}
Taking the logarithm also avoids extreme behavior when the denominator terms are small. Despite the hole illuminations that can be time-varying, $A_{pqrs}$ is expected to be fixed in time except for measurement noise. Therefore, it allows us to average the closure amplitude measurements over time, before or after taking the logarithm. 

As before, this takes the form 
\begin{align}
    \mathbf{y} &= \mathbf{J} \mathbf{x} + \mathbf{n} \, ,
\end{align}
where, 
$\mathbf{x} \coloneqq \begin{bmatrix}
    a & b & c
\end{bmatrix}^\intercal$, 
$\mathbf{J}$ is the \textit{Jacobian} in Eq.~(\ref{eqn:logClAmp}) consisting of 3 columns that transforms $\mathbf{x}$ to $\mathbf{y} \coloneqq \begin{bmatrix}
    \langle log A_{0123}\rangle & \ldots & \langle \log A_{2340}\rangle
\end{bmatrix}^\intercal$, and
$\mathbf{n}$ is some random noise vector associated with the time-averaged measurements, $\mathbf{y}$. 

The number of quads possible from $N_\textrm{h}$ holes is $\mathcal{O}(N_\textrm{h}^4)$. However, only $N_\textrm{h}(N_\textrm{h}-3)/2$ carry independent information, from which the rest can be determined \cite{TMS2017}. We follow the method presented in \cite{Blackburn+2020} to determine the independent quads. For $N_\textrm{h}=5$, we obtain $N_\textrm{meas}=N_\textrm{h}(N_\textrm{h}-3)/2=5$ measurements from independent quads from which $N_\textrm{par}=3$ beam shape parameters are to be determined. Because $N_\textrm{par}\le N_\textrm{meas}$ for $N_\textrm{h}\geq 5$, a unique solution in the form, $\widetilde{\mathbf{x}}=\mathbf{W}\mathbf{y}$, is possible using Eq.~(\ref{eqn:solution}). If only $N_\textrm{h}=4$ non-redundant holes used, there will be only $N_\textrm{meas}=2$ independent closure amplitude measurements which will be inadequate to constrain the $N_\textrm{par}=3$ Gaussian beam shape parameters. This is consistent with the earlier reasoning that $N_\textrm{h}=5$ is the minimum number of non-redundant holes required for an overdetermined system of equations. Increasing the number of non-redundant holes further can be verified to comfortably satisfy this criterion.

\begin{figure}
\includegraphics[width=\linewidth]
{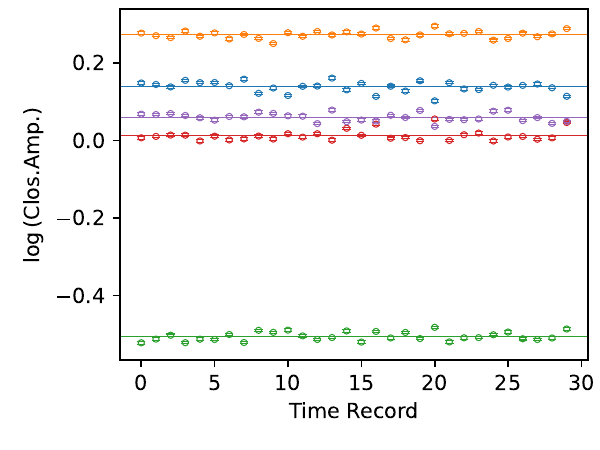}
\caption{Logarithm of measured closure amplitudes of 5 independent quads of individual time records (unfilled circles) and averaged over time (lines). Error bars, shown on the measurements, are smaller than the symbol size due to the high light levels.
\label{fig:logClAmp-measurements}}
\end{figure}

Fig.~\ref{fig:logClAmp-measurements} shows the measurements of the closure amplitudes from independent quads after taking the logarithm. Empty circles denote measurements at individual instances of time, while the lines denote the average across time. The measurement errors are typically smaller than the symbol sizes due to the high light levels. 

The parameters and the covariance between them are obtained from Eqs.~(\ref{eqn:solution}) and (\ref{eqn:out-covariance}), respectively. Thus, the beam shape can be estimated independent of the effects of aperture illumination in contrast to the reference methods, LASC or NLASC. Another notable difference in this method is that because the hole illuminations are immaterial (time-varying or not), the measurements can be averaged together over a period of time to improve the signal-to-noise ratio, from which the solution needs to be estimated only once resulting in a reduced data volume and a smaller computational footprint at a slower cadence. The shape parameters so determined are converted to the Gaussian parameters using Eqs.~(\ref{eqn:semi-principal-axes}) and (\ref{eqn:pa}). 

\section{Results}\label{sec:results}

\begin{figure}
\includegraphics[width=\linewidth]
{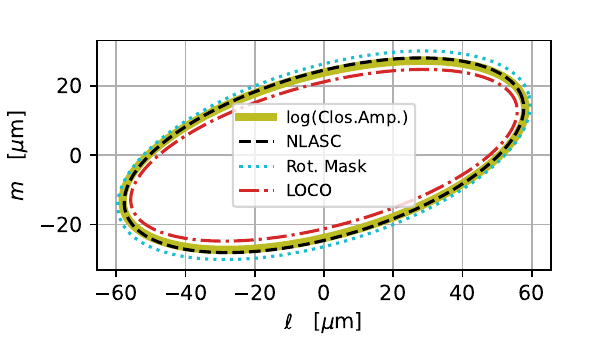}
\caption{Comparison of elliptical Gaussian shapes for the synchrotron beam estimated from the closure amplitude approach in this paper (solid green), NLASC (dashed black), rotated 2-hole mask method (dotted blue), and LOCO (dot-dashed red).
\label{fig:beam-shape}}
\end{figure}

The estimates, with uncertainties, for the Gaussian shape parameters-- the demonstrative result from our method using closure amplitudes-- are: 
\begin{itemize}
    \item semi-major axis, $\sigma_\textrm{maj} = 59.93\substack{+0.07 \\ -0.07}$~$\mu$m, 
    \item semi-minor axis, $\sigma_\textrm{min} = 23.19\substack{+0.50 \\ -0.52}$~$\mu$m, and 
    \item beam tilt, $\theta = 15\fdg 20\substack{+0.16 \\ -0.16}$.
\end{itemize}
Fig.~\ref{fig:beam-shape} compares the elliptical Gaussian-shaped beam profile determined from our closure amplitude method against the various methods obtained from the same experimental setup, namely NLASC \cite{Nikolic+2024}, 2-hole mask rotation \cite{Torino+Iriso2016}, and Linear Optics from Closed Orbits (LOCO; \cite{Safranek1997}). The results between methods are consistent to within a few percent. 

\begin{figure}
\includegraphics[width=\linewidth]
{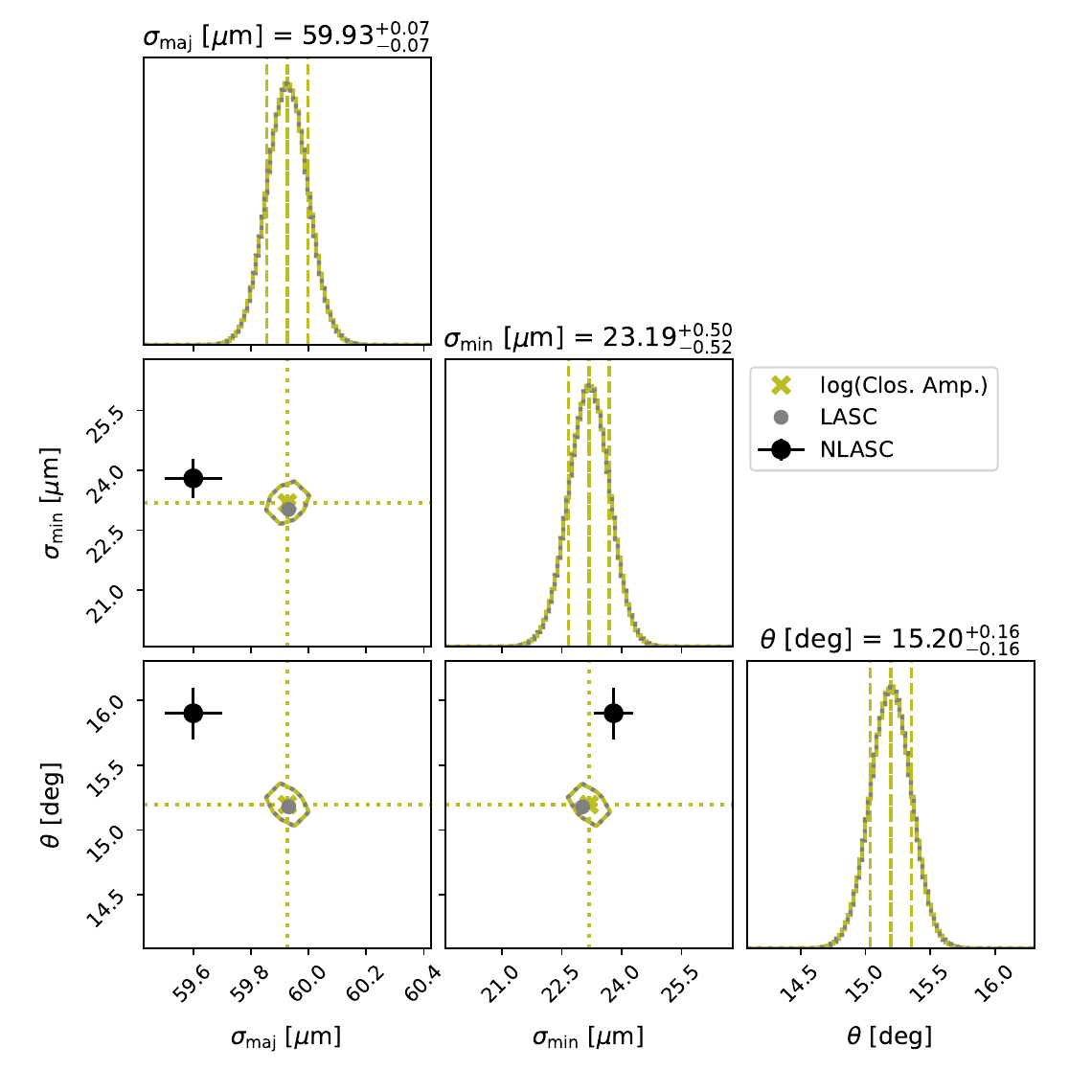}
\caption{Estimates of elliptical Gaussian beam shape parameters, the covariance between their uncertainties, and marginalized distributions from the closure amplitude approach in this paper (green) and the reference LASC method (gray circle and dotted gray lines). The contours correspond to $1\sigma$-equivalent of a two-dimensional Gaussian probability distribution (enclosing 39.3\% of the distribution's volume). Dotted green lines show the means, while the three vertical dashed green lines in the marginalized one-dimensional distributions show the median and the inner 68\% (also listed above the histograms). Estimates from NLASC (black) are overlaid. 
\label{fig:logClAmp-covariance}}
\end{figure}

To characterize the uncertainties in these estimates, we show the covariances between the parameters in Fig.~\ref{fig:logClAmp-covariance}. The solution, $\widetilde{\mathbf{x}}$, denoting estimates of $a$, $b$, and $c$, and the covariance between them, $\langle\bm{\varepsilon\varepsilon}^\intercal\rangle$, derived using our closure amplitudes method were used to draw a sufficiently large number of multivariate normal samples and converted to Gaussian shape parameters using Eqs.~(\ref{eqn:semi-principal-axes}) and (\ref{eqn:pa}). Their distributions are shown as covariant contours and marginalized histograms in solid green. For comparison, a similar covariance plot obtained from averaging the frame-by-frame solution of the reference method, LASC, are overlaid in gray. The two distributions are virtually identical to each other. In addition, the estimates from another reference method, the NLASC, are also shown in black. Although the results from the closure amplitudes and LASC methods exhibit a statistical difference relative to that of the NLASC, the semi-major and semi-minor axes agree to within one percent, and the beam tilt to within one degree. This difference is likely due to differences in the mathematical treatments between the methods. For example, NLASC includes the auto-correlations (total intensities through the holes) whereas we only use the visibilities corresponding to cross-correlations in the closure amplitude and LASC methods. Also, our inverse covariance weighting, though mathematically optimal, relies on covariance determined empirically from the data which could be affected by a few bad frames, whereas the NLASC does not use the same weighting scheme. These differences will be investigated with larger data sets in future. The hole illuminations estimated from the two reference methods, LASC and NLASC, were verified to be in very good agreement with each other (not shown here). 

Note that the choice of independent quads is not unique. We tried other independent combinations and verified the results (optimum estimate, $\widetilde{\mathbf{x}}$, and the covariance, $\langle\bm{\varepsilon\,\varepsilon}^\intercal\rangle$) to be identical. This confirms that any complete set of independent quads contains identical information as any other, and the use of inverse covariance weighting ensures an optimum solution\footnote{This may be because we are in a high signal-to-noise regime. \cite{Nakajima+1989,Kulkarni1989} studied the statistical independence of bispectrum measurements and concluded that at low light levels, all triads contain useful information and must be used. A similar argument may apply for closure amplitude statistics at low light levels.}.

\section{Conclusions}\label{sec:conclusions}

We introduce a novel method using closure amplitudes, a concept well-established in radio interferometry, for directly estimating the transverse two-dimensional shape of a light beam. The key advantage of this approach is its immunity to non-uniform hole illumination and the errors therein, unlike other multi-hole interferometry methods that require accurate calibration of hole illuminations, separately or simultaneously. Closure amplitudes inherently bypass the need for such calibration, simplifying the estimation process by directly inferring the two-dimensional transverse profile of the light beam.

Because closure amplitudes depend only on the intrinsic transverse beam shape, they can be averaged over time without being affected by variations in the hole illuminations, provided the beam parameters remain stable. This leads to efficient, reliable measurements at a practical cadence afforded by the averaging over longer time intervals, lower data rates, and reduced computational overhead.
We have validated the method of directly determining the two-dimensional beam shape using closure amplitudes with data from the ALBA synchrotron light source, demonstrating consistency with existing techniques.

This approach to light beam shape determination represents a novel adaption of the powerful concept of closure amplitudes from radio to optical interferometry. Our results show that this methodology can be effectively extended to multi-element interferometry at optical and other wavelengths. 

{\bf Acknowledgments.} The National Radio Astronomy Observatory is a facility of the National Science Foundation operated under cooperative agreement by Associated Universities, Inc. 
Related patents and patent applications: Patent No. US 12,104,901 B2; Provisional Patent No. No. 63/355,174; Provisional Patent No. 63/648,303; US Application No. 19/079,876.

{\bf Disclosure Statement:} The authors declare no conflicts of interest.

{\bf Data availability:} The ALBA interferograms  underlying the results presented in this paper are not publicly available at this time but may be obtained from the authors upon reasonable request.

\bibliography{refs}


\end{document}